
%
%
\expandafter\ifx\csname phyzzx\endcsname\relax
 \message{It is better to use PHYZZX format than to
          \string\input\space PHYZZX}\else
 \wlog{PHYZZX macros are already loaded and are not
          \string\input\space again}%
 \endinput \fi
\catcode`\@=11 
\let\rel@x=\relax
\let\n@expand=\relax
\def\pr@tect{\let\n@expand=\noexpand}
\let\protect=\pr@tect
\let\gl@bal=\global
%
%
%
\newfam\cpfam
\newdimen\b@gheight             \b@gheight=12pt
\newcount\f@ntkey               \f@ntkey=0
\def\f@m{\afterassignment\samef@nt\f@ntkey=}
\def\samef@nt{\fam=\f@ntkey \the\textfont\f@ntkey\rel@x}
\def\setstr@t{\setbox\strutbox=\hbox{\vrule height 0.85\b@gheight
                                depth 0.35\b@gheight width\z@ }}
\input phyzzx.fonts
%
\def\rm{\n@expand\f@m0 }
\def\mit{\n@expand\f@m1 }         
\def\cal{\n@expand\f@m2 }
\def\it{\n@expand\f@m\itfam}
\def\sl{\n@expand\f@m\slfam}
\def\bf{\n@expand\f@m\bffam}
\def\tt{\n@expand\f@m\ttfam}
\def\caps{\n@expand\f@m\cpfam}    
\def\em@{\rel@x\ifnum\f@ntkey=0 \it \else
        \ifnum\f@ntkey=\bffam \it \else \rm \fi \fi }
\def\em{\n@expand\em@}
\def\fourteenpoint{\fourteenf@nts \samef@nt \b@gheight=14pt \setstr@t }
\def\twelvepoint{\twelvef@nts \samef@nt \b@gheight=12pt \setstr@t }
\def\tenpoint{\tenf@nts \samef@nt \b@gheight=10pt \setstr@t }
\normalbaselineskip = 20pt plus 0.2pt minus 0.1pt
\normallineskip = 1.5pt plus 0.1pt minus 0.1pt
\normallineskiplimit = 1.5pt
\newskip\normaldisplayskip
\normaldisplayskip = 20pt plus 5pt minus 10pt
\newskip\normaldispshortskip
\normaldispshortskip = 6pt plus 5pt
\newskip\normalparskip
\normalparskip = 6pt plus 2pt minus 1pt
\newskip\skipregister
\skipregister = 5pt plus 2pt minus 1.5pt
\newif\ifsingl@
\newif\ifdoubl@
\newif\iftwelv@  \twelv@true
\def\singlespace{\singl@true\doubl@false\spaces@t}
\def\doublespace{\singl@false\doubl@true\spaces@t}
\def\normalspace{\singl@false\doubl@false\spaces@t}
\def\Tenpoint{\tenpoint\twelv@false\spaces@t}
\def\Twelvepoint{\twelvepoint\twelv@true\spaces@t}
\def\spaces@t{\rel@x
      \iftwelv@ \ifsingl@\subspaces@t3:4;\else\subspaces@t1:1;\fi
       \else \ifsingl@\subspaces@t3:5;\else\subspaces@t4:5;\fi \fi
      \ifdoubl@ \multiply\baselineskip by 5
         \divide\baselineskip by 4 \fi }
\def\subspaces@t#1:#2;{
      \baselineskip = \normalbaselineskip
      \multiply\baselineskip by #1 \divide\baselineskip by #2
      \lineskip = \normallineskip
      \multiply\lineskip by #1 \divide\lineskip by #2
      \lineskiplimit = \normallineskiplimit
      \multiply\lineskiplimit by #1 \divide\lineskiplimit by #2
      \parskip = \normalparskip
      \multiply\parskip by #1 \divide\parskip by #2
      \abovedisplayskip = \normaldisplayskip
      \multiply\abovedisplayskip by #1 \divide\abovedisplayskip by #2
      \belowdisplayskip = \abovedisplayskip
      \abovedisplayshortskip = \normaldispshortskip
      \multiply\abovedisplayshortskip by #1
        \divide\abovedisplayshortskip by #2
      \belowdisplayshortskip = \abovedisplayshortskip
      \advance\belowdisplayshortskip by \belowdisplayskip
      \divide\belowdisplayshortskip by 2
      \smallskipamount = \skipregister
      \multiply\smallskipamount by #1 \divide\smallskipamount by #2
      \medskipamount = \smallskipamount \multiply\medskipamount by 2
      \bigskipamount = \smallskipamount \multiply\bigskipamount by 4 }
\def\normalbaselines{ \baselineskip=\normalbaselineskip
   \lineskip=\normallineskip \lineskiplimit=\normallineskip
   \iftwelv@\else \multiply\baselineskip by 4 \divide\baselineskip by 5
     \multiply\lineskiplimit by 4 \divide\lineskiplimit by 5
     \multiply\lineskip by 4 \divide\lineskip by 5 \fi }
\Twelvepoint  
\interlinepenalty=50
\interfootnotelinepenalty=5000
\predisplaypenalty=9000
\postdisplaypenalty=500
\hfuzz=1pt
\vfuzz=0.2pt
\newdimen\HOFFSET  \HOFFSET=0pt
\newdimen\VOFFSET  \VOFFSET=0pt
\newdimen\HSWING   \HSWING=0pt
\dimen\footins=8in
%
%
%
\newskip\pagebottomfiller
\pagebottomfiller=\z@ plus \z@ minus \z@
\def\pagecontents{
   \ifvoid\topins\else\unvbox\topins\vskip\skip\topins\fi
   \dimen@ = \dp255 \unvbox255
   \vskip\pagebottomfiller
   \ifvoid\footins\else\vskip\skip\footins\footrule\unvbox\footins\fi
   \ifr@ggedbottom \kern-\dimen@ \vfil \fi }
\def\makeheadline{\vbox to 0pt{ \skip@=\topskip
      \advance\skip@ by -12pt \advance\skip@ by -2\normalbaselineskip
      \vskip\skip@ \line{\vbox to 12pt{}\the\headline} \vss
      }\nointerlineskip}
\def\makefootline{\baselineskip = 1.5\normalbaselineskip
                 \line{\the\footline}}
\newif\iffrontpage
\newif\ifp@genum
\def\nopagenumbers{\p@genumfalse}
\def\pagenumbers{\p@genumtrue}
\pagenumbers
\newtoks\paperheadline
\newtoks\paperfootline
\newtoks\letterheadline
\newtoks\letterfootline
\newtoks\letterinfo
\newtoks\date
\paperheadline={\hfil}
\paperfootline={\hss\iffrontpage\else\ifp@genum\tenrm\folio\hss\fi\fi}
\letterheadline{\iffrontpage \hfil \else
    \rm \ifp@genum page~~\folio\fi \hfil\the\date \fi}
\letterfootline={\iffrontpage\the\letterinfo\else\hfil\fi}
\letterinfo={\hfil}
\def\monthname{\rel@x\ifcase\month 0/\or January\or February\or
   March\or April\or May\or June\or July\or August\or September\or
   October\or November\or December\else\number\month/\fi}
\def\today{\monthname~\number\day, \number\year}
\date={\today}
\headline=\paperheadline 
\footline=\paperfootline 
\countdef\pageno=1      \countdef\pagen@=0
\countdef\pagenumber=1  \pagenumber=1
\def\advancepageno{\gl@bal\advance\pagen@ by 1
   \ifnum\pagenumber<0 \gl@bal\advance\pagenumber by -1
    \else\gl@bal\advance\pagenumber by 1 \fi
    \gl@bal\frontpagefalse  \swing@ }
\def\folio{\ifnum\pagenumber<0 \romannumeral-\pagenumber
           \else \number\pagenumber \fi }
\def\swing@{\ifodd\pagenumber \gl@bal\advance\hoffset by -\HSWING
             \else \gl@bal\advance\hoffset by \HSWING \fi }
\def\footrule{\dimen@=\prevdepth\nointerlineskip
   \vbox to 0pt{\vskip -0.25\baselineskip \hrule width 0.35\hsize \vss}
   \prevdepth=\dimen@ }
\let\footnotespecial=\rel@x
\newdimen\footindent
\footindent=24pt
\def\Textindent#1{\noindent\llap{#1\enspace}\ignorespaces}
\def\Vfootnote#1{\insert\footins\bgroup
   \interlinepenalty=\interfootnotelinepenalty \floatingpenalty=20000
   \singl@true\doubl@false\Tenpoint
   \splittopskip=\ht\strutbox \boxmaxdepth=\dp\strutbox
   \leftskip=\footindent \rightskip=\z@skip
   \parindent=0.5\footindent \parfillskip=0pt plus 1fil
   \spaceskip=\z@skip \xspaceskip=\z@skip \footnotespecial
   \Textindent{#1}\footstrut\futurelet\next\fo@t}

\def\vfootnote#1{\Vfootnote{${#1}$}}
\def\footnote#1{\attach{#1}\vfootnote{#1}}

\let\footsymbol=\star
\newcount\lastf@@t           \lastf@@t=-1
\newcount\footsymbolcount    \footsymbolcount=0
\newif\ifPhysRev
\def\bumpfootsymbolcount{\rel@x
   \iffrontpage \bumpfootsymbolpos \else \advance\lastf@@t by 1
     \ifPhysRev \bumpfootsymbolneg \else \bumpfootsymbolpos \fi \fi
   \gl@bal\lastf@@t=\pagen@ }
\def\bumpfootsymbolpos{\ifnum\footsymbolcount <0
                            \gl@bal\footsymbolcount =0 \fi
    \ifnum\lastf@@t<\pagen@ \gl@bal\footsymbolcount=0
     \else \gl@bal\advance\footsymbolcount by 1 \fi }
\def\bumpfootsymbolneg{\ifnum\footsymbolcount >0
             \gl@bal\footsymbolcount =0 \fi
         \gl@bal\advance\footsymbolcount by -1 }
\def\fd@f#1 {\xdef\footsymbol{\mathchar"#1 }}
\def\generatefootsymbol{\ifcase\footsymbolcount \fd@f 13F \or \fd@f 279
        \or \fd@f 27A \or \fd@f 278 \or \fd@f 27B \else
        \ifnum\footsymbolcount <0 \fd@f{023 \number-\footsymbolcount }
         \else \fd@f 203 {\loop \ifnum\footsymbolcount >5
                \fd@f{203 \footsymbol } \advance\footsymbolcount by -1
                \repeat }\fi \fi }

\def\nonfrenchspacing{\sfcode`\.=3001 \sfcode`\!=3000 \sfcode`\?=3000
        \sfcode`\:=2000 \sfcode`\;=1500 \sfcode`\,=1251 }
\nonfrenchspacing
\newdimen\d@twidth
{\setbox0=\hbox{s.} \gl@bal\d@twidth=\wd0 \setbox0=\hbox{s}
        \gl@bal\advance\d@twidth by -\wd0 }
\def\removehglue{\loop \unskip \ifdim\lastskip >\z@ \repeat }
\def\roll@ver#1{\removehglue \nobreak \count255 =\spacefactor \dimen@=\z@
        \ifnum\count255 =3001 \dimen@=\d@twidth \fi
        \ifnum\count255 =1251 \dimen@=\d@twidth \fi
    \iftwelv@ \kern-\dimen@ \else \kern-0.83\dimen@ \fi
   #1\spacefactor=\count255 }
\def\step@ver#1{\rel@x \ifmmode #1\else \ifhmode
        \roll@ver{${}#1$}\else {\setbox0=\hbox{${}#1$}}\fi\fi }
\def\attach#1{\step@ver{\strut^{\mkern 2mu #1} }}
%
%
%
\newcount\chapternumber      \chapternumber=0
\newcount\sectionnumber      \sectionnumber=0
\newcount\equanumber         \equanumber=0
\let\chapterlabel=\rel@x
\let\sectionlabel=\rel@x
\newtoks\chapterstyle        \chapterstyle={\Number}
\newtoks\sectionstyle        \sectionstyle={\chapterlabel.\Number}
\newskip\chapterskip         \chapterskip=\bigskipamount
\newskip\sectionskip         \sectionskip=\medskipamount
\newskip\headskip            \headskip=8pt plus 3pt minus 3pt
\newdimen\chapterminspace    \chapterminspace=15pc
\newdimen\sectionminspace    \sectionminspace=10pc
\newdimen\referenceminspace  \referenceminspace=20pc
\def\chapterreset{\gl@bal\advance\chapternumber by 1
   \ifnum\equanumber<0 \else\gl@bal\equanumber=0\fi
   \sectionnumber=0 \let\sectionlabel=\rel@x
   {\pr@tect\xdef\chapterlabel{\the\chapterstyle{\the\chapternumber}}}}
\def\alphabetic#1{\count255='140 \advance\count255 by #1\char\count255}
\def\Alphabetic#1{\count255='100 \advance\count255 by #1\char\count255}
\def\Roman#1{\uppercase\expandafter{\romannumeral #1}}
\def\roman#1{\romannumeral #1}
\def\Number#1{\number #1}
\def\BLANC#1{}
\def\titleparagraphs{\interlinepenalty=9999
     \leftskip=0.03\hsize plus 0.22\hsize minus 0.03\hsize
     \rightskip=\leftskip \parfillskip=0pt
     \hyphenpenalty=9000 \exhyphenpenalty=9000
     \tolerance=9999 \pretolerance=9000
     \spaceskip=0.333em \xspaceskip=0.5em }
\def\titlestyle#1{\par\begingroup \titleparagraphs
     \iftwelv@\fourteenpoint\else\twelvepoint\fi
   \noindent #1\par\endgroup }
\def\spacecheck#1{\dimen@=\pagegoal\advance\dimen@ by -\pagetotal
   \ifdim\dimen@<#1 \ifdim\dimen@>0pt \vfil\break \fi\fi}
\def\chapter#1{\par \penalty-300 \vskip\chapterskip
   \spacecheck\chapterminspace
   \chapterreset \titlestyle{\chapterlabel.~#1}
   \nobreak\vskip\headskip \penalty 30000
   {\pr@tect\wlog{\string\chapter\space \chapterlabel}} }

\def\section#1{\par \ifnum\the\lastpenalty=30000\else
   \penalty-200\vskip\sectionskip \spacecheck\sectionminspace\fi
   \gl@bal\advance\sectionnumber by 1
   {\pr@tect
   \xdef\sectionlabel{\the\sectionstyle\the\sectionnumber}
   \wlog{\string\section\space \sectionlabel}}
   \noindent {\caps\enspace\sectionlabel.~~#1}\par
   \nobreak\vskip\headskip \penalty 30000 }
\def\subsection#1{\par
   \ifnum\the\lastpenalty=30000\else \penalty-100\smallskip \fi
   \noindent\undertext{#1}\enspace \vadjust{\penalty5000}}

\def\undertext#1{\vtop{\hbox{#1}\kern 1pt \hrule}}
\def\APPENDIX#1#2{\par\penalty-300\vskip\chapterskip
   \spacecheck\chapterminspace \chapterreset \xdef\chapterlabel{#1}
   \titlestyle{APPENDIX #2} \nobreak\vskip\headskip \penalty 30000
   \wlog{\string\Appendix~\chapterlabel} }
\def\Appendix#1{\APPENDIX{#1}{#1}}
\def\appendix{\APPENDIX{A}{}}
\def\unnumberedchapters{\let\makechapterlabel=\rel@x
      \let\chapterlabel=\rel@x  \sectionstyle={\BLANC}
      \let\sectionlabel=\rel@x \sequentialequations }
%
%
%
\def\eqname#1{\rel@x {\pr@tect
  \ifnum\equanumber<0 \xdef#1{{\rm(\number-\equanumber)}}%
     \gl@bal\advance\equanumber by -1
  \else \gl@bal\advance\equanumber by 1
     \ifx\chapterlabel\rel@x \def\d@t{}\else \def\d@t{.}\fi
    \xdef#1{{\rm(\chapterlabel\d@t\number\equanumber)}}\fi #1}}
\def\eqinsert#1{\noalign{\dimen@=\prevdepth \nointerlineskip
   \setbox0=\hbox to\displaywidth{\hfil #1}
   \vbox to 0pt{\kern 0.5\baselineskip\hbox{$\!\box0\!$}\vss}
   \prevdepth=\dimen@}}
%

%
%
\def\GENITEM#1;#2{\par \hangafter=0 \hangindent=#1
    \Textindent{$ #2 $}\ignorespaces}
\outer\def\newitem#1=#2;{\gdef#1{\GENITEM #2;}}

\newdimen\itemsize                \itemsize=30pt
\newitem\item=1\itemsize;
\newitem\sitem=1.75\itemsize;     
\newitem\ssitem=2.5\itemsize;     
\outer\def\newlist#1=#2&#3&#4;{\toks0={#2}\toks1={#3}%
   \count255=\escapechar \escapechar=-1
   \alloc@0\list\countdef\insc@unt\listcount     \listcount=0
   \edef#1{\par
      \countdef\listcount=\the\allocationnumber
      \advance\listcount by 1
      \hangafter=0 \hangindent=#4
      \Textindent{\the\toks0{\listcount}\the\toks1}}
   \expandafter\expandafter\expandafter
    \edef\c@t#1{begin}{\par
      \countdef\listcount=\the\allocationnumber \listcount=1
      \hangafter=0 \hangindent=#4
      \Textindent{\the\toks0{\listcount}\the\toks1}}
   \expandafter\expandafter\expandafter
    \edef\c@t#1{con}{\par \hangafter=0 \hangindent=#4 \noindent}
   \escapechar=\count255}
\def\c@t#1#2{\csname\string#1#2\endcsname}
\newlist\point=\Number&.&1.0\itemsize;
\newlist\subpoint=(\alphabetic&)&1.75\itemsize;
\newlist\subsubpoint=(\roman&)&2.5\itemsize;
%

%
%
%
%
\newcount\referencecount     \referencecount=0
\newcount\lastrefsbegincount \lastrefsbegincount=0
\newif\ifreferenceopen       \newwrite\referencewrite
\newdimen\refindent          \refindent=30pt
\def\normalrefmark#1{\attach{\scriptscriptstyle [ #1 ] }}
\let\PRrefmark=\attach
\def\NPrefmark#1{\step@ver{{\;[#1]}}}
\def\refmark#1{\rel@x\ifPhysRev\PRrefmark{#1}\else\normalrefmark{#1}\fi}
\def\refend@{\refmark{\number\referencecount}}
\def\refend{\refend@{}\space }
\def\refsend{\refmark{\count255=\referencecount
   \advance\count255 by-\lastrefsbegincount
   \ifcase\count255 \number\referencecount
   \or \number\lastrefsbegincount,\number\referencecount
   \else \number\lastrefsbegincount-\number\referencecount \fi}\space }
\def\REFNUM#1{\rel@x \gl@bal\advance\referencecount by 1
    \xdef#1{\the\referencecount }}
\def\Refnum#1{\REFNUM #1\refend@ } 
\def\REF#1{\REFNUM #1\R@FWRITE\ignorespaces}
\def\Ref#1{\Refnum #1\REFWRITE }
\def\ref{\Ref\?}
\def\REFS#1{\REFNUM #1\gl@bal\lastrefsbegincount=\referencecount
    \REFWRITE }

\def\r@fitem#1{\par \hangafter=0 \hangindent=\refindent \Textindent{#1}}
\def\refitem#1{\r@fitem{#1.}}
\def\NPrefitem#1{\r@fitem{[#1]}}
\def\NPrefs{\let\refmark=\NPrefmark \let\refitem=NPrefitem}
\def\REFWRITE{\R@FWRITE\rel@x }
\def\R@FWRITE#1{\ifreferenceopen \else \gl@bal\referenceopentrue
     \immediate\openout\referencewrite=\jobname.refs
     \toks@={\begingroup \refoutspecials \catcode`\^^M=10 }%
     \immediate\write\referencewrite{\the\toks@}\fi
    \immediate\write\referencewrite{\noexpand\refitem %
                                    {\the\referencecount}}%
    \p@rse@ndwrite \referencewrite #1}
\begingroup
 \catcode`\^^M=\active \let^^M=\relax %
 \gdef\p@rse@ndwrite#1#2{\begingroup \catcode`\^^M=12 \newlinechar=`\^^M%
         \chardef\rw@write=#1\sc@nlines#2}%
 \gdef\sc@nlines#1#2{\sc@n@line \g@rbage #2^^M\endsc@n \endgroup #1}%
 \gdef\sc@n@line#1^^M{\expandafter\toks@\expandafter{\deg@rbage #1}%
         \immediate\write\rw@write{\the\toks@}%
         \futurelet\n@xt \sc@ntest }%
\endgroup
\def\sc@ntest{\ifx\n@xt\endsc@n \let\n@xt=\rel@x
       \else \let\n@xt=\sc@n@notherline \fi \n@xt }
\def\sc@n@notherline{\sc@n@line \g@rbage }
\def\deg@rbage#1{}
\let\g@rbage=\relax    \let\endsc@n=\relax
\def\refout{\par\penalty-400\vskip\chapterskip
   \spacecheck\referenceminspace
   \ifreferenceopen \Closeout\referencewrite \referenceopenfalse \fi
   \line{\fourteenrm\hfil REFERENCES\hfil}\vskip\headskip
   \input \jobname.refs
   }
\def\refoutspecials{\sfcode`\.=1000 \interlinepenalty=1000
         \rightskip=\z@ plus 1em minus \z@ }
\def\Closeout#1{\toks0={\par\endgroup}\immediate\write#1{\the\toks0}%
   \immediate\closeout#1}
%
%
\newcount\figurecount     \figurecount=0
\newcount\tablecount      \tablecount=0
\newif\iffigureopen       \newwrite\figurewrite
\newif\iftableopen        \newwrite\tablewrite
\def\FIGNUM#1{\rel@x \gl@bal\advance\figurecount by 1
    \xdef#1{\the\figurecount}}
\def\FIGURE#1{\FIGNUM #1\F@GWRITE\ignorespaces }

\def\figitem#1{\r@fitem{#1)}}
\def\FIGWRITE{\F@GWRITE\rel@x }
\def\TABNUM#1{\rel@x \gl@bal\advance\tablecount by 1
    \xdef#1{\the\tablecount}}
\def\TABLE#1{\TABNUM #1\T@BWRITE\ignorespaces }

\def\tabitem#1{\r@fitem{#1:}}
\def\TABWRITE{\T@BWRITE\rel@x }
\def\F@GWRITE#1{\iffigureopen \else \gl@bal\figureopentrue
     \immediate\openout\figurewrite=\jobname.figs
     \toks@={\begingroup \catcode`\^^M=10 }%
     \immediate\write\figurewrite{\the\toks@}\fi
    \immediate\write\figurewrite{\noexpand\figitem %
                                 {\the\figurecount}}%
    \p@rse@ndwrite \figurewrite #1}
\def\T@BWRITE#1{\iftableopen \else \gl@bal\tableopentrue
     \immediate\openout\tablewrite=\jobname.tabs
     \toks@={\begingroup \catcode`\^^M=10 }%
     \immediate\write\tablewrite{\the\toks@}\fi
    \immediate\write\tablewrite{\noexpand\tabitem %
                                 {\the\tablecount}}%
    \p@rse@ndwrite \tablewrite #1}
\def\figout{\par\penalty-400
   \vskip\chapterskip\spacecheck\referenceminspace
   \iffigureopen \Closeout\figurewrite \figureopenfalse \fi
   \line{\fourteenrm\hfil FIGURE CAPTIONS\hfil}\vskip\headskip
   \input \jobname.figs
   }
\def\tabout{\par\penalty-400
   \vskip\chapterskip\spacecheck\referenceminspace
   \iftableopen \Closeout\tablewrite \tableopenfalse \fi
   \line{\fourteenrm\hfil TABLE CAPTIONS\hfil}\vskip\headskip
   \input \jobname.tabs
   }
%
%
%
\newbox\picturebox
\def\p@cht{\ht\picturebox }
\def\p@cwd{\wd\picturebox }
\def\p@cdp{\dp\picturebox }
\newdimen\xshift
\newdimen\yshift
\newdimen\captionwidth
\newskip\captionskip
\captionskip=15pt plus 5pt minus 3pt
\def\fullwidth{\captionwidth=\hsize }
\newtoks\Caption
\newif\ifcaptioned
\newif\ifselfcaptioned
\def\caption{\captionedtrue \Caption }
\newcount\linesabove
\newif\iffileexists
\newtoks\picfilename
\def\fil@#1 {\fileexiststrue \picfilename={#1}}
\def\file#1{\if=#1\let\n@xt=\fil@ \else \def\n@xt{\fil@ #1}\fi \n@xt }
\def\pl@t{\begingroup \pr@tect
    \setbox\picturebox=\hbox{}\fileexistsfalse
    \let\height=\p@cht \let\width=\p@cwd \let\depth=\p@cdp
    \xshift=\z@ \yshift=\z@ \captionwidth=\z@
    \Caption={}\captionedfalse
    \linesabove =0 \picturedefault }
\def\plot{\pl@t \selfcaptionedfalse }
\def\Picture#1{\gl@bal\advance\figurecount by 1
    \xdef#1{\the\figurecount}\pl@t \selfcaptionedtrue }

\def\s@vepicture{\iffileexists \parsefilename \redopicturebox \fi
   \ifdim\captionwidth>\z@ \else \captionwidth=\p@cwd \fi
   \xdef\lastpicture{\iffileexists
        \setbox0=\hbox{\raise\the\yshift \vbox{%
              \moveright\the\xshift\hbox{\picturedefinition}}}%
        \else \setbox0=\hbox{}\fi
         \ht0=\the\p@cht \wd0=\the\p@cwd \dp0=\the\p@cdp
         \vbox{\hsize=\the\captionwidth \line{\hss\box0 \hss }%
              \ifcaptioned \vskip\the\captionskip \noexpand\Tenpoint
                \ifselfcaptioned Figure~\the\figurecount.\enspace \fi
                \the\Caption \fi }}%
    \endgroup }
\let\endpicture=\s@vepicture
\def\savepicture#1{\s@vepicture \global\let#1=\lastpicture }
\def\displaypicture{\fullwidth \s@vepicture $$\lastpicture $${}}
\def\toppicture{\fullwidth \s@vepicture \topinsert
    \lastpicture \medskip \endinsert }
\def\midpicture{\fullwidth \s@vepicture \midinsert
    \lastpicture \endinsert }
%
%
\def\leftpicture{\pres@tpicture
    \dimen@i=\hsize \advance\dimen@i by -\dimen@ii
    \setbox\picturebox=\hbox to \hsize {\box0 \hss }%
    \wr@paround }
\def\rightpicture{\pres@tpicture
    \dimen@i=\z@
    \setbox\picturebox=\hbox to \hsize {\hss \box0 }%
    \wr@paround }
\def\pres@tpicture{\gl@bal\linesabove=\linesabove
    \s@vepicture \setbox\picturebox=\vbox{
         \kern \linesabove\baselineskip \kern 0.3\baselineskip
         \lastpicture \kern 0.3\baselineskip }%
    \dimen@=\p@cht \dimen@i=\dimen@
    \advance\dimen@i by \pagetotal
    \par \ifdim\dimen@i>\pagegoal \vfil\break \fi
    \dimen@ii=\hsize
    \advance\dimen@ii by -\parindent \advance\dimen@ii by -\p@cwd
    \setbox0=\vbox to\z@{\kern-\baselineskip \unvbox\picturebox \vss }}
\def\wr@paround{\Caption={}\count255=1
    \loop \ifnum \linesabove >0
         \advance\linesabove by -1 \advance\count255 by 1
         \advance\dimen@ by -\baselineskip
         \expandafter\Caption \expandafter{\the\Caption \z@ \hsize }%
      \repeat
    \loop \ifdim \dimen@ >\z@
         \advance\count255 by 1 \advance\dimen@ by -\baselineskip
         \expandafter\Caption \expandafter{%
             \the\Caption \dimen@i \dimen@ii }%
      \repeat
    \edef\n@xt{\parshape=\the\count255 \the\Caption \z@ \hsize }%
    \par\noindent \n@xt \strut \vadjust{\box\picturebox }}
\let\picturedefault=\relax
\let\parsefilename=\relax
\def\redopicturebox{\let\picturedefinition=\rel@x
   \errhelp=\disabledpictures
   \errmessage{This version of TeX cannot handle pictures.  Sorry.}}
\newhelp\disabledpictures
     {You will get a blank box in place of your picture.}
%
%
%
%
%
%
%
%
%
%
\def\FRONTPAGE{\ifvoid255\else\vfill\penalty-20000\fi
   \gl@bal\pagenumber=1     \gl@bal\chapternumber=0
   \gl@bal\equanumber=0     \gl@bal\sectionnumber=0
   \gl@bal\referencecount=0 \gl@bal\figurecount=0
   \gl@bal\tablecount=0     \gl@bal\frontpagetrue
   \gl@bal\lastf@@t=0       \gl@bal\footsymbolcount=0}

\def\papers{\papersize\headline=\paperheadline\footline=\paperfootline}
\def\papersize{\hsize=35pc \vsize=50pc \hoffset=0pc \voffset=1pc
   \advance\hoffset by\HOFFSET \advance\voffset by\VOFFSET
   \pagebottomfiller=0pc
   \skip\footins=\bigskipamount \normalspace }
\papers  
%
%
\newskip\lettertopskip       \lettertopskip=20pt plus 50pt
\newskip\letterbottomskip    \letterbottomskip=\z@ plus 100pt
\newskip\signatureskip       \signatureskip=40pt plus 3pt
\def\lettersize{\hsize=6.5in \vsize=8.5in \hoffset=0in \voffset=0.5in
   \advance\hoffset by\HOFFSET \advance\voffset by\VOFFSET
   \pagebottomfiller=\letterbottomskip
   \skip\footins=\smallskipamount \multiply\skip\footins by 3
   \singlespace }
\def\MEMO{\lettersize \headline=\letterheadline \footline={\hfil }%
   \let\rule=\memorule \FRONTPAGE \memohead }

\def\memodate{\afterassignment\MEMO \date }
\def\memit@m#1{\smallskip \hangafter=0 \hangindent=1in
    \Textindent{\caps #1}}
\def\subject{\memit@m{Subject:}}
\def\topic{\memit@m{Topic:}}
\def\from{\memit@m{From:}}
\def\to{\rel@x \ifmmode \rightarrow \else \memit@m{To:}\fi }
\def\memorule{\medskip\hrule height 1pt\bigskip}  
\def\memohead{\centerline{\fourteenrm MEMORANDUM}}
\newwrite\labelswrite
\newtoks\rw@toks
\def\letters{\lettersize
   \headline=\letterheadline \footline=\letterfootline
   \immediate\openout\labelswrite=\jobname.lab}

\let\letterhead=\rel@x
\def\addressee#1{\medskip\line{\hskip 0.75\hsize plus\z@ minus 0.25\hsize
                               \the\date \hfil }%
   \vskip \lettertopskip
   \ialign to\hsize{\strut ##\hfil\tabskip 0pt plus \hsize \crcr #1\crcr}
   \writelabel{#1}\medskip \noindent\hskip -\spaceskip \ignorespaces }
\def\rwl@begin#1\cr{\rw@toks={#1\crcr}\rel@x
   \immediate\write\labelswrite{\the\rw@toks}\futurelet\n@xt\rwl@next}
\def\rwl@next{\ifx\n@xt\rwl@end \let\n@xt=\rel@x
      \else \let\n@xt=\rwl@begin \fi \n@xt}
\let\rwl@end=\rel@x
\def\writelabel#1{\immediate\write\labelswrite{\noexpand\labelbegin}
     \rwl@begin #1\cr\rwl@end
     \immediate\write\labelswrite{\noexpand\labelend}}
\newtoks\FromAddress         \FromAddress={}
\newtoks\sendername          \sendername={}
\newbox\FromLabelBox
\newdimen\labelwidth          \labelwidth=6in
\def\makelabels{\afterassignment\Makelabels \sendersname=}
\def\Makelabels{\FRONTPAGE \letterinfo={\hfil } \MakeFromBox
     \immediate\closeout\labelswrite  \input \jobname.lab\vfil\eject}
\let\labelend=\rel@x
\def\labelbegin#1\labelend{\setbox0=\vbox{\ialign{##\hfil\cr #1\crcr}}
     \MakeALabel }
\def\MakeFromBox{\gl@bal\setbox\FromLabelBox=\vbox{\Tenpoint
     \ialign{##\hfil\cr \the\sendername \the\FromAddress \crcr }}}
\def\MakeALabel{\vskip 1pt \hbox{\vrule \vbox{
        \hsize=\labelwidth \hrule\bigskip
        \leftline{\hskip 1\parindent \copy\FromLabelBox}\bigskip
        \centerline{\hfil \box0 } \bigskip \hrule
        }\vrule } \vskip 1pt plus 1fil }
\def\signed#1{\par \nobreak \bigskip \dt@pfalse \begingroup
  \everycr={\noalign{\nobreak
            \ifdt@p\vskip\signatureskip\gl@bal\dt@pfalse\fi }}%
  \tabskip=0.5\hsize plus \z@ minus 0.5\hsize
  \halign to\hsize {\strut ##\hfil\tabskip=\z@ plus 1fil minus \z@\crcr
          \noalign{\gl@bal\dt@ptrue}#1\crcr }%
  \endgroup \bigskip }
\newbox\letterb@x
\def\lettertext{\par \vskip\parskip \unvcopy\letterb@x \par }
\def\multiletter{\setbox\letterb@x=\vbox\bgroup
      \everypar{\vrule height 1\baselineskip depth 0pt width 0pt }
      \singlespace \topskip=\baselineskip }
\def\letterend{\par\egroup}
%
%
%
\newskip\frontpageskip
\newtoks\Pubnum   
\newtoks\Pubtype  \let\pubtype=\Pubtype
\newif\ifp@bblock  \p@bblocktrue
\def\PH@SR@V{\doubl@true \baselineskip=24.1pt plus 0.2pt minus 0.1pt
             \parskip= 3pt plus 2pt minus 1pt }
\def\PHYSREV{\papers\PhysRevtrue\PH@SR@V}

\def\titlepage{\FRONTPAGE\papers\ifPhysRev\PH@SR@V\fi
   \ifp@bblock\p@bblock \else\hrule height\z@ \rel@x \fi }
\def\nopubblock{\p@bblockfalse}

\frontpageskip=12pt plus .5fil minus 2pt
\Pubtype={}
\Pubnum={}
\def\p@bblock{\begingroup \tabskip=\hsize minus \hsize
   \baselineskip=1.5\ht\strutbox \topspace-2\baselineskip
   \halign to\hsize{\strut ##\hfil\tabskip=0pt\crcr
       \the\Pubnum\crcr\the\date\crcr\the\pubtype\crcr}\endgroup}
\def\title#1{\vskip\frontpageskip \titlestyle{#1} \vskip\headskip }
\def\author#1{\vskip\frontpageskip\titlestyle{\twelvecp #1}\nobreak}

\def\address#1{\par\kern 5pt\titlestyle{\twelvepoint\it #1}}
\def\andaddress{\par\kern 5pt \centerline{\sl and} \address}

\def\abstract{\par\dimen@=\prevdepth \hrule height\z@ \prevdepth=\dimen@
   \vskip\frontpageskip\centerline{\fourteenrm ABSTRACT}\vskip\headskip }

%
%
%

\def\\{\rel@x \ifmmode \backslash \else {\tt\char`\\}\fi }
\def\sequentialequations{\rel@x \if\equanumber<0 \else
  \gl@bal\equanumber=-\equanumber \gl@bal\advance\equanumber by -1 \fi }
\def\nextline{\unskip\nobreak\hfill\break}

\def\journal#1&#2(#3){\begingroup \let\journal=\dummyj@urnal
    \unskip, \sl #1\unskip~\bf\ignorespaces #2\rm
    (\afterassignment\j@ur \count255=#3), \endgroup\ignorespaces }
\def\j@ur{\ifnum\count255<100 \advance\count255 by 1900 \fi
          \number\count255 }
\def\dummyj@urnal{%
    \toks@={Reference foul up: nested \journal macros}%
    \errhelp={Your forgot & or ( ) after the last \journal}%
    \errmessage{\the\toks@ }}

\def\topspace{\hrule height 0pt depth 0pt \vskip}

\def\Buildrel#1\under#2{\mathrel{\mathop{#2}\limits_{#1}}}
\def\becomes#1{\mathchoice{\becomes@\scriptstyle{#1}}
   {\becomes@\scriptstyle{#1}} {\becomes@\scriptscriptstyle{#1}}
   {\becomes@\scriptscriptstyle{#1}}}
\def\becomes@#1#2{\mathrel{\setbox0=\hbox{$\m@th #1{\,#2\,}$}%
        \mathop{\hbox to \wd0 {\rightarrowfill}}\limits_{#2}}}

\let\int=\intop         
\def\lsim{\mathrel{\mathpalette\@versim<}}
\def\gsim{\mathrel{\mathpalette\@versim>}}
\def\@versim#1#2{\vcenter{\offinterlineskip
        \ialign{$\m@th#1\hfil##\hfil$\crcr#2\crcr\sim\crcr } }}
\def\big#1{{\hbox{$\left#1\vbox to 0.85\b@gheight{}\right.\n@space$}}}
\def\Big#1{{\hbox{$\left#1\vbox to 1.15\b@gheight{}\right.\n@space$}}}
\def\bigg#1{{\hbox{$\left#1\vbox to 1.45\b@gheight{}\right.\n@space$}}}
\def\Bigg#1{{\hbox{$\left#1\vbox to 1.75\b@gheight{}\right.\n@space$}}}
\def\){\mskip 2mu\nobreak }
%
%
%
\let\sec@nt=\sec
\def\sec{\rel@x\ifmmode\let\n@xt=\sec@nt\else\let\n@xt\section\fi\n@xt}
\def\obsolete#1{\message{Macro \string #1 is obsolete.}}
\def\firstsec#1{\obsolete\firstsec \section{#1}}
\def\firstsubsec#1{\obsolete\firstsubsec \subsection{#1}}
\def\thispage#1{\obsolete\thispage \gl@bal\pagenumber=#1\frontpagefalse}
\def\thischapter#1{\obsolete\thischapter \gl@bal\chapternumber=#1}
\def\splitout{\obsolete\splitout\rel@x}
\def\prop{\obsolete\prop \propto }
\def\nextequation#1{\obsolete\nextequation \gl@bal\equanumber=#1
   \ifnum\the\equanumber>0 \gl@bal\advance\equanumber by 1 \fi}
\def\BOXITEM{\afterassigment\B@XITEM\setbox0=}
\def\B@XITEM{\par\hangindent\wd0 \noindent\box0 }
%
%
%
\nopagenumbers

\centerline{\bf LONG-LIVED CHARGED HEAVY LEPTONS}
\vskip 0.5cm
\centerline{MARC SHER}
\vskip 0.2cm
\centerline{Physics Dept., Coll. of William and Mary,
Williamsburg VA  23187}\centerline{sher@wmheg.physics.wm.edu}
\vskip 0.5cm

Since the accurate measurement of the parameters of $Z^o$ decay,
it has been known that there exist only three light neutrinos
coupling to the $Z^o$ in the manner prescribed by the standard model.
However, this does not rule out the possibility that additional
lepton families exist in which the masses of both the neutrino and
the charged lepton exceed half the mass of the $Z^o$.  Such heavy
leptons, which will be designated as $L$ and $N$, for the charged and
neutral varieties respectively, will be a target of investigation at
the next generation of particle colliders.

Many analyses of heavy lepton production have previously been done[1].
In this talk[2], I will discuss a possibility that has not been a focus
of previous investigation--namely, the possibility that the $N$ is
heavier than (or extremely close in mass to) the $L$.  In the
simplest extension of the standard model, with a fourth generation
lepton left-handed doublet and right-handed $N$ and $L$ singlets,
the masses are arbitrary, and thus in half of parameter-space, the
$N$ is heavier.  In another simple extension, with a vectorlike
lepton family, the masses are degenerate at tree-level.  Thus, it is
plausible that $N$ is heavier than or close in mass to the $L$.  In
this case, the $L$ can only decay via mixing with lighter
generations, and thus could be very long-lived.  This, as we will
see, has major implications for the phenomenological detection of
such particles.

Two cases will be discussed in the talk.  In model I, which is the
simplest extension of the standard model which incorporates very massive
neutrinos, the right-handed $N$ and $L$ are singlets, and we will assume that
the $N$ is heavier.  The $L$ can then only decay (primarily into $\nu\ +\
W$) via mixing with lighter generations.   For a $100$ GeV $L$, the lifetime
will be given by (in seconds) $10^{-21}/\sin^2\theta$, where $\theta$ is the
mixing angle.  In model II, the right-handed $N$ and $L$ form an isodoublet,
and their masses are degenerate at tree-level.  In this case, radiative
corrections cause a mass-splitting of $O(250)$ MeV, leading to a lifetime
given by the smaller of $10^{-21}/\sin^2\theta$ and $10^{-9\pm 2}$ seconds.
Clearly, if the mixing angle is very small, then the $L$ can be very
long-lived.

Before discussing the plausible values of the mixing angle, we
first consider the possibility that the mixing angle is extremely
small or zero, and that the $L$ has a long enough lifetime to have
cosmological implications (model I).  This possibility has been
extensively analyzed in Ref. 3.  The $L$'s cannot be stable, since their
abundance would be sufficiently large that the $L^-$'s would have
combined with protons, making "heavy hydrogen", and these would have
been detected in searches for anomalously heavy hydrogen in
sea-water.  The $L$'s would also have been detected directly in
plastic track detectors.  Thus, they must decay with a lifetime
less than the age of the universe.  Since photons will be
emitted in the decay (since the $L$ is charged), these photons could
be detected directly.  Failure to observe such photons force the $L$
to have a lifetime less than $10^{13}$ seconds.  Finally, if the
lifetime is greater than $10^9$ seconds ($100$ years), the photons
would disrupt the cosmic microwave background, leading to distortions
that would have been seen by COBE.  Thus, the lifetime of the $L$
must be less than about $100$ years.  See [3] for details.

If the mixing angle is nonzero, then what are plausible values?  In
see-saw type models, $\sin^2\theta$ is given by either $m_\tau/m_L$
of $m_{\nu_\tau}/m_N$, depending on whether the mixing can occur in
the charged lepton sector or is confined to the neutrino sector.  In
the former case, the lifetime is $O(10^{-19})$ seconds; i.e. the $L$
will decay at the vertex.  In the latter case, however, the lifetime
is $O(10^{-11}{\rm\ sec})(10\ {\rm eV}/m_{\nu_\tau})$.  Thus the
lifetime is at least a picosecond, and could easily be long enough to
pass completely through any detector.

One can see that all three lifetimes: (a) decay at the vertex, (b)
decay in the detector and (c) decay outside the detector are possible.
Let us consider each possibility in turn.  If the decay is at the
vertex, the standard heavy lepton analysis will apply, and it will
appear as if the $L$ decays into a massless neutrino and a real
or virtual $W$.  In Model II (in which the $L$ and $N$ form a
vectorlike doublet), however, there is a GIM violation which leads to
the decay $L\rightarrow \tau\ Z$.  The branching ratio for this mode
is a complicated function of (unknown) mixing angles, but typically
varies between $1$ and $100$ percent.  Even at $1$ percent, the
background for the $\tau Z$ mode will be significantly smaller than
the $\nu_\tau W$ mode, and thus the GIM violating mode may be the
easiest method of detection.

If the decay is in the detector, but away from the vertex, it should
be easy to detect.  An apparent muon will suddenly decay into missing
energy and a real or virtual $W$.  The backgrounds should be
negligible.

If the decay is outside the detector, the $L$ will be
indistinguishable from a muon.  The production cross-section (see [2]
for a detailed analysis) is large enough that thousands of $L$'s
would be produced annually at the SSC, but is small compared with the
muon pair production cross-section, and thus the ``extra" muons
wouldn't be noticed.  One possible method of detection would be
time-of-flight.  Many of the $L$'s will have $\beta<0.9$ (see [2] for
specific graphs), and thus if timing information is available, the
$L$'s could be seen.  It is interesting that thousands of $L$'s could
be produced, but that they would be missed if timing information is
not present.

Finally, one can consider the case in which (in Model I) the mixing
angle is generated by Planck scale effects (see [4] for several such
models in which an angle of this size occurs naturally).  In this
case, the lifetime turns out to be approximately $10$ years, barely
 under the
cosmological bound.  In this case, one can seriously consider
stopping and collecting the $L$'s.  In Ref. 4, it is shown that several
thousand a year could be produced, stopped and collected in an $e^+e^-$
collider with an energy of roughly ten percent above threshold, surrounded
by a segmented detector of liquid argon.  They could then be transported away
from the detector site.  What could they be used for?  The number produced
is to small to use as an energy storage device or explosive, the mass is
too heavy to be practical in fusion catalysis (the limiting factor is the
time between a fusion and capture by the next proton--we estimate under a
watt of energy per $L$ could be produced). One could learn a great deal
about nuclear structure from the capture process, although perhaps not much
more than one can learn from other electromagnetic probes (such as CEBAF).
Obviously, it is premature to speculate in much detail as to the uses of
such heavy leptons.

 Very little is known about the mass spectrum of fermions in
the standard model.  In this talk, I have pointed out that in the simplest
extension of the standard model, in which all right-handed fields are
singlets, it is quite possible that a fourth-generation neutrino will be
heavier that its charged counterpart.  The latter will then be very
long-lived.  Three plausible values of the mixing angle give very different
phenomenology; the charged lepton will either decay at the vertex (albeit
into a massless neutrino), away from the vertex but within the detector, or
will pass completely through the detector.  In the latter case, then if the
detectors at the LHC and SSC have the ability to determine the velocity of
these particles, they will look just like muons and be missed.  In the
extreme case in which the mixing angle is only generated by Planck scale
effects (it can't be zero for cosmological reasons), not only is detection
possible, but collection for further study may be possible.  Finally, in an
alternative model in which the fourth generation leptons form an
isodoublet, the possible decay of the charged lepton into $\tau Z$ offers a
promising, nearly background free, method of detection.\parindent=0pt
\vskip 0.5 cm
\centerline{\bf REFERENCES}
 1.\ R.M. Barnett, et al., Proceedings of the Summer Study on High
Energy Physics in the 1990s, ed. S. Jensen, World Scientific (1988) 157.
\nextline
2.\ P. Frampton, D. Ng, M. Sher and Y. Yuan, Phys Rev. {\bf D48}
(1993) 3128.
\nextline 3.\ M. Sher and Y. Yuan, Phys. Lett. {\bf B285} (1992) 336.
\nextline
4.\ J. Goity, W.J. Kossler and M. Sher, Phys. Rev. {\bf
 D48} (1993) xxxx.
\end